\documentclass[journal,10pt,onecolumn,draftclsnofoot,]{IEEEtran}

\usepackage{ifpdf}
\usepackage{cite}

\ifCLASSINFOpdf
   \usepackage[pdftex]{graphicx}
   \graphicspath{{/figures/}}
   \DeclareGraphicsExtensions{.pdf,.jpeg,.png}
\else
\fi
\usepackage{amsmath}
\usepackage{{amssymb}}
\usepackage{array}

\usepackage{url}
\usepackage{subfigure}          % for subfigures

\begin{document}
\title{Decentralized Control of Large Collaborative Swarms using Random Finite Set Theory}

\author{Bryce~Doerr,~\IEEEmembership{Member,~IEEE,}
        and~Richard~Linares,~\IEEEmembership{Member,~IEEE}
        % <-this % stops a space
        \thanks{Manuscript received Dec. 17, 2019;}
\thanks{Bryce Doerr is with the Department of Aeronautics and Astronautics, Massachusetts Institute of Technology, Cambridge, MA, USA (email: bdoerr@mit.edu)}% <-this % stops a space
\thanks{Richard Linares is with the Department of Aeronautics and Astronautics, Massachusetts Institute of Technology, Cambridge, MA, USA, (email: linaresr@mit.edu)}% <-this % stops a space
}

\maketitle

\begin{abstract}
Controlling large swarms of robotic agents presents many challenges including, but not limited to, computational complexity due to a large number of agents, uncertainty in the functionality of each agent in the swarm, and uncertainty in the swarm's configuration. The contribution of this work is to decentralize Random Finite Set (RFS) control of large collaborative swarms for control of individual agents. The RFS control formulation assumes that the topology underlying the swarm control is complete and uses the complete graph in a centralized manner. To generalize the control topology in a localized or decentralized manner, sparse LQR is used to sparsify the RFS control gain matrix obtained using iterative LQR. This allows agents to use information of agents near each other (localized topology) or only the agent's own information (decentralized topology) to make a control decision. Sparsity and performance for decentralized RFS control are compared for different degrees of localization in feedback control gains which show that the stability and performance compared to centralized control do not degrade significantly in providing RFS control for large collaborative swarms.
\end{abstract}


% \fi
\IEEEpeerreviewmaketitle

\section{Introduction}
\IEEEPARstart{C}{ontrol} of large collaborative networks or swarms is currently an emerging area in controls research. A swarm network is typically comprised of tiny robots programmed with limited actuators that perform specific tasks in the network formation. For example, the swarm can use its combined effort to grasp or move in the environment which can offer a better way to meet a goal compared to the abilities of a single agent \cite{1}. Specifically in space applications, swarm control of satellites and rovers can be used for the exploration of asteroids and other celestial bodies of interest \cite{2} or areas of assembly and construction on-orbit including constructing space observatories and space habitats \cite{izzo2005mission}. Swarms involving UAVs have proven to be widely useful in military applications such as search and rescue missions, communication relaying, border patrol, surveillance, and mapping of hostile territory \cite{3}. From these engineering applications, the use of collaborative swarms is an attractive option to meet objectives that may be too difficult for a single agent.

For collaborative swarms, several control techniques have been implemented to date. With centralized control, one agent in the swarm computes the overall swarm control and manages the control execution for individual agents allowing it to oversee the other agents' system processes \cite{5}. Unfortunately, centralized control suffers from two main problems. As the number of agents in the swarm increases, the computational workload becomes more expensive \cite{bakule2008decentralized}. This is especially true when the swarm agents are low-cost and are located in an unknown environment. Additionally, centralized control is not robust against individual agent failures \cite{6}. With a thousand low-cost agents present in a swarm, communication, actuation, and sensing are performed with less reliability. Thus, centralized control may not be a viable option for these systems.  

Changing how the model for the representation and behavior for a swarm state in space and time has been shown to alleviate the computational complexity of control methods and solutions \cite{37,38,39,41,45}. Previously, the swarm/potential model using the random finite set (RFS) formalism was used to describe the temporal evolution of the probabilistic description of the robotic swarm to promote decentralized coordination \cite{28}. By using a measure-value recursion of the RFS formalism for the swarm agents, the swarm dynamics can be determined with computational efficiency.

\begin{figure}
\begin{centering}
      \includegraphics[keepaspectratio,trim={0 .55cm 0 .35cm},clip,width=.48\textwidth]{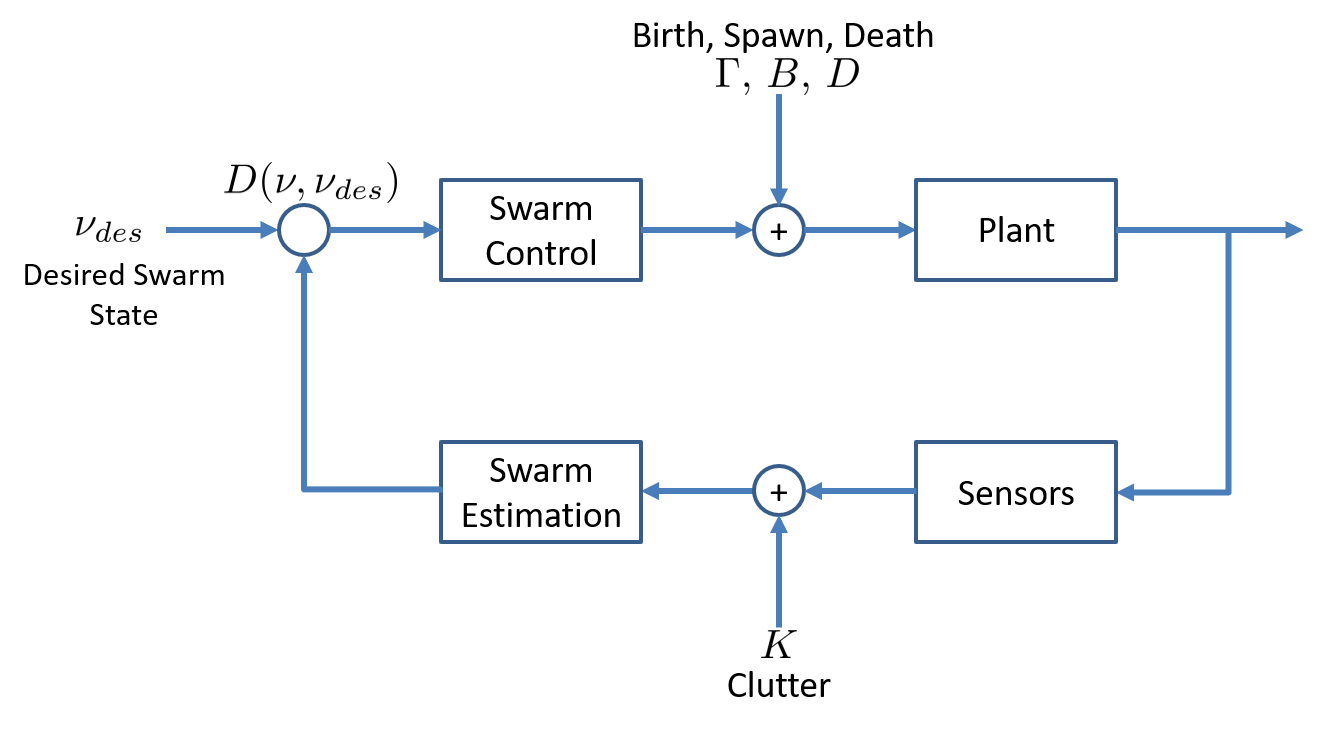}
        \caption{A block diagram of the RFS control and estimation architecture in a closed-loop. }\label{closedloop}
\end{centering}
\end{figure}

This RFS formulation was then expanded for control of large collaborative swarms \cite{doerr2018control}. This work generalized the state representation of the control problem as an RFS, where an RFS is a collection of agent states, with no ordering between individual agents, that can randomly change through time \cite{mahler2003multitarget}. Figure \ref{closedloop} shows the concept of the contributed work, where the first moment of the RFS is used as the state, $\nu$, and the desired RFS swarm configuration is defined by its first moment, $\nu_{des}$. The novelty of this work was to generalize the notion of distance using RFS-based distance measures and to ``close-the-loop" by processing measurements from an unknown number of agents with defined spawn ($B$), birth ($\Gamma$), and death ($D$) rates to obtain a multi-agent estimate for control using the Gaussian mixture Probability Hypothesis Density (GM-PHD) filter and a variant of differential dynamic programming (DDP) called iterative linear quadratic regulator (ILQR). In this example, the topology underlying the swarm control is complete and uses the complete graph in a centralized manner. To obtain a complete graph for RFS control, the swarm is estimated in both cardinality (number of agents) and state using the GM-PHD filter. RFS control through ILQR approximates a quadratic value function from the distributional distance-based cost, and it is iterated to determine an optimal control solution. The results combining the PHD filter and ILQR using the RFS formalism provided implicit proof for RFS control of large collaborative swarms.

Although the RFS formalism allows for varying swarm states and number of agents with time, a central low-cost agent with limited computational capacity may have difficulty computing a centralized control command due to the large number of agents. Thus, it is necessary to break down the centralized control problem into smaller, more manageable subproblems which are weakly dependent or independent from each other. This becomes the area of decentralized or localized control. Decentralized control is able to control agents in a swarm by using different techniques on the swarm control (information) structure. Two different methods are of interest for decentralized control. The first area is the development of decentralized controllers under specific structural constraints \cite{fardad2009optimal,jovanovic2010optimality,fardad2011design,lin2011augmented}. An example of a structural constraint is sparsity requirements for an agent in the swarm which suggests that it only has access to the information structure from agents near it. The other area of interest is the development of decentralized control under communication constraints (delays). By adding delay and uncertainty into multi-agent systems, control can be degraded. Convex methods and optimal control have been tools used to develop decentralized systems that incorporate communication delays \cite{voulgaris2003optimal,bamieh2005convex}.

In the original RFS control problem, the control (information) topology is assumed to be complete using all the state information obtained from the GM-PHD filter. This is centralized control in which the swarm computes the overall swarm control and manages the control execution for individual agents, allowing it to oversee the other agents' control processes.

For decentralized RFS control, the control topology is used in a localized or decentralized manner using sparse control matrices. The decentralized RFS control is realized using sparse LQR to sparsify the centralized RFS control gain matrix obtained using ILQR. This allows agents to use local information topology (information of agents near each other) or a fully decentralized topology (information of the agent's own information) to make a control decision. Sparse LQR allows for more stability and less performance degradation than truncating a centralized control matrix may provide. Sparsity and performance for decentralized RFS control are compared for different degrees of localization in the feedback control gains which show the viability for decentralized control for large collaborative swarms.

\section{RFS Formulation}
The RFS formulation was first considered in \cite{mahler2003multitarget} and implemented in a tractable recursion for multi-target estimation in \cite{vo2006gaussian}. Then, the tractable recursion was used in conjunction with an RFS formulation for control of large collaborative swarms in \cite{doerr2018control}. For a complete background of RFS control and estimation, \cite{vo2006gaussian,doerr2018control,doerr2019random} are referred. The discussion below provides the neccessary background for forming decentralized RFS control.

The multi-agent problem considers the Bayesian recursion through an RFS formulation with discrete-time dynamics \cite{vo2006gaussian}. This theory addresses the decentralized formulation for each agent in the formation. Each agent has the challenge of estimating its local formation configuration and designing a control policy to achieve some local configuration. It is assumed that each agent within the swarm is identical and that using unique identifiers on each agent is unnecessary. Using RFS theory, the number of agents and their states is determined from measurements. The agents in the field may die, survive and move into the next state through dynamics, or appear by spawning or birthing.  The number of agents in the field is denoted by $N_{\text{total}}(t)$ and may be randomly varying at each time-step by the union of the birth ($\Gamma_k:\emptyset\rightarrow \left\lbrace {\bf x}_{i,k},{\bf x}_{i+1,k},\cdots,{\bf x}_{i+N_{birth},k}\right\rbrace$), spawn ($B_{k|k-1}\left( \zeta\right):{\bf x}_{i,k-1}\rightarrow \left\lbrace {\bf x}_{i,k},{\bf x}_{i+1,k},\cdots,{\bf x}_{i+N_{spawn},k}\right\rbrace$), and surviving ($S_{k|k-1}\left( \zeta\right):{\bf x}_{i,k-1}\rightarrow{\bf x}_{i,k}$) agents. Death is denoted by $D_k\left( \zeta\right):{\bf x}_{i,k-1}\rightarrow\emptyset$. Note that ${\bf x}_{i,k}$ is for the $i$th swarm agent's state. This is described by an RFS, $X_k$, given by
\begin{equation}\label{unionstate}
X_k=\left[ \bigcup_{\zeta\in X_{k-1}} S_{k|k-1}\left( \zeta\right)\right]\cup\left[ \bigcup_{\zeta\in X_{k-1}}B_{k|k-1}\left( \zeta\right)\right]\cup \Gamma_k.
\end{equation}
$X_k= \left\{{\bf x}_{1,k},{\bf x}_{2,k},\cdots,{\bf x}_{N_{total(k)},k} \right\}$ denotes a realization of the RFS distribution for agents. The individual RFSs in Eq. \eqref{unionstate} are assumed to be independent from each other. For example, any births that occur at any time-step are independent from any surviving agents.
%Additionally, the control of each unit is represented by an RFS, $U_t= \left\{{\bf u}_{1,t},{\bf u}_{2,t},\cdots,{\bf u}_{N_{total(t)},t} \right\}$.  
At any time, $k$, the RFS probability density function can be written as
\begin{equation}\label{rfsstate}
\begin{split}
p(X_k&=\left\{{\bf x}_{1,k},{\bf x}_{2,k},\dots,{\bf x}_{n,k}\right\})\\&=p(|X_k|=n)\times p(\left\{{\bf x}_{1,k},{\bf x}_{2,k},\cdots,{\bf x}_{n,k}\right\}{\mid \lvert X_k\rvert =n)}.
\end{split}
\end{equation}
For a generalized observation process, the agents are either detected ($\Theta_k\left( {\bf x}_k\right):{\bf x}_{i,k}\rightarrow{\bf y}_{i,k}$), or not detected ($F_k\left( {\bf x}_k\right):{\bf x}_{i,k}\rightarrow\emptyset$). Clutter or false alarms ($K_k: \emptyset\rightarrow \left\lbrace {\bf y}_{1,k},{\bf y}_{2,k},\cdots,{\bf y}_{N_{clutter},k}\right\rbrace$), defined as measurements that do not belong to any agents, are also present in the set of observations. Note that ${\bf y}_{i,k}$ is for the $i$th swarm agent's measurement. Therefore, the RFS of measurements is described by
\begin{equation}\label{unionmeasurement}
Z_k=K_k\cup\left[ \bigcup_{ {\bf x_k}\in X_k }\Theta_k\left(  {\bf x}_k\right)\right],
\end{equation}
where the origins of each measurement are not known and unique identifiers are not necessary. Again, the individual RFSs in Eq. \eqref{unionmeasurement} are independent of each other, so measurements and clutter are obtained independently from each other. Single-agent filtering cannot be applied because measurements cannot be associated with the agent that generated it. By using the RFS formulation, measurements can be associated to individual agents in the swarm.

The control sequence is also defined by an RFS in the form $U_k= \left\{{\bf u}_{1,k},{\bf u}_{2,k},\dots,{\bf u}_{N_{total(k)},k} \right\}$ and an RFS probability density in a form similar to Eq. \eqref{rfsstate} since the number of the agents in the field to be controlled is also varying.

The RFS formulation of describing multi-agent states ($X_k$) and observations ($Z_k$) can be described by a state transition, $f_{k|k-1}\left( X_{k}|X_{k-1}\right)$, and a measurement likelihood, $g_k\left( Z_{k}|X_{k}\right)$, function. To determine the multi-agent posterior density, a multi-agent Bayes recursion is used given by
\begin{subequations}
\begin{equation}
\begin{split}
p_{k|k-1}\left( X_{k}|Z_{1:k-1}\right)&=\int f_{k|k-1}\left( X_{k}|X_{k-1}\right)\\&\times p_{k-1}\left( X_{k-1}|Z_{1:k-1}\right)\mu_s(dX_{k-1}),
\end{split}
\end{equation}
\begin{equation}
p_{k}\left( X_{k}|Z_{1:k}\right)=\frac{g_k\left( Z_{k}|X_{k}\right)p_{k|k-1}\left( X_{k}|Z_{1:k-1}\right)}{\int g_k\left( Z_{k}|X_{k}\right)p_{k|k-1}\left( X_{k}|Z_{1:k-1}\right)\mu_s(dX_{k})},
\end{equation}
\end{subequations}
where $\mu_s$ is a reference measure on some function $F(X)$. The recursion is computationally expensive due to multiple integrals, but solutions have been found for a small number of targets using sequential Monte Carlo \cite{35}. Fortunately, a PHD filter approximation provides computational tractability for large number of agents.
\subsection{Probability Hypothesis Density (PHD) Filter}
Instead of propagating the multi-agent posterior density through a multi-agent Bayes recursion, the Probability Hypothesis Density (PHD) filter propagates the posterior intensity function. The nonnegative intensity function, $v({\bf x})$, is a first-order statistical moment of the RFS state that represents the probability of finding an agent in a region of state space $S$. The expected number of agents in the region $S$ is the integral of the intensity function given by
\begin{equation}
\mathbb{E}(|X\cap S|)=\int |X\cap S|P(dX)=\int_Sv({\bf x})d{\bf x},
\end{equation}
where the expectation represents an RFS $X$ intersecting a region $S$ with a probability distribution $P$ dependent on $X$. This gives the total mass or the expected number of agents of RFS $X$ in a region $S$. The local maximum in intensity $v({\bf x})$ shows the highest concentration of expected number of agents which can be used to determine an estimate for the agents in $X$ at a time-step.

Poisson RFS are fully characterized by their intensities. By assuming the RFS $X$ is Poisson of the form $p(|X|=n)$ and $p(\left\{{\bf x}_1,{\bf x}_2,...,{\bf x}_n\right\}{\mid \lvert X\rvert =n)}$, approximate solutions can be determined by the PHD filter. Propagation of the PHD can be determined if the agents are assumed to be independent and identically distributed with the cardinality of the agent set that is Poisson distributed \cite{vo2006gaussian}. Secondly, it is assumed that the agents' motion and measurements are independent of each other. Thirdly, clutter and birth RFSs are assumed to be Poisson RFSs and clutter is independent from the measurement RFS. Lastly, the time-update multi-target density $p_{k|k-1}$ is Poisson, but if there is no spawning and the surviving and birth RFSs are Poisson, then this assumption is satisfied. It is noted that the assumptions made by the PHD filter are strong assumptions for swarming robotics. However, this is a good starting point for an initial proof-of-concept study. 
The PHD recursion for a general intensity function, $v_t({\bf x})$, is given by
\begin{subequations}\label{rfsphdfilter}
\begin{equation}\label{rfs1}
\bar{v}_t({\bf x})=b({\bf x})+\int p_s(\zeta)f({\bf x}|\zeta)v(\zeta)d\zeta+\int \beta({\bf x}|\zeta)v(\zeta)d\zeta,
\end{equation}
where $b({\bf x})$, $p_s(\zeta)$, and $\beta({\bf x}|\zeta)$  are the agents' birth, survival, and spawn intensity, $f({\bf x}|\zeta)$ is the target motion model, and $\zeta$ is the previous state respectively \cite{vo2006gaussian}. The bar on $\bar{v}_t({\bf x})$ denotes that the PHD has been time-updated.
%conditioned on measurements $Z_{t-1}$.
For the measurement update, the equation is given by

\begin{equation} \label{rfs2}
\resizebox{0.48\textwidth}{!}{$
{v_t({\bf x})=(1-p_d({\bf x}))\bar{v}_t({\bf x})}+\sum_{{\bf z}\in Z_t}\frac{p_d({\bf x})g({\bf z}_t|{\bf x})\bar{v}_t({\bf x})}{c({\bf z})+\int p_d(\zeta)g({\bf z}_t|\zeta)\bar{v}_t(\zeta)d\zeta},$}
\end{equation}
\end{subequations}
where $p_d({\bf x})$, $g({\bf z}_t|{\bf x})$, and $c({\bf z})$ are the probability of detection, likelihood function, and clutter model of the sensor respectively \cite{vo2006gaussian}. By using this recursion, the swarm probabilistic description can be updated. The recursion itself avoids computations that arise from the unknown relation between agents and its measurements, and the posterior intensity is a function of a single agent's state space. Unfortunately, Eq. \eqref{rfsphdfilter} does not contain a closed-form solution and the numerical integration suffers from higher computational time as the state increases due to an increasing number of agents. 
\subsection{Gaussian Mixture Model and Control Formulation}
Fortunately, a closed-form solution exists if it is assumed that the survival and detection probabilities are state independent (i.e. $p_s({\bf x})=p_s$ and $p_d({\bf x})=p_d$), and the intensities of the birth and spawn RFSs are Gaussian mixtures.
To formulate the optimal control problem, the current and desired intensities are
\begin{equation}\label{funcff}
\begin{split}
\bar{\nu}({\bf x},k)\triangleq\sum_{i=1}^{N_f}w_f^{(i)}\mathcal{N}\left({\bf x};{\bf m}_f^i,P_f^i\right)&\triangleq\nu_b({\bf x},k)+\nu_{p_s}({\bf x},k)\\&+\nu_\beta({\bf x},k),
\end{split}
\end{equation}
\begin{equation}\label{funcg}
\nu_{des}({\bf x},k)\triangleq g({\bf x})\triangleq\sum_{i=1}^{N_g}w_g^{(i)}\mathcal{N}\left({{\bf x};\bf m}_g^i,P_g^i\right),
\end{equation}
where $w^{(i)}$ are the weights and $\mathcal{N}\left({\bf x};{\bf m}^i,P^i\right)$ is the probability density function of an $i$th multivariate Gaussian distribution with a mean and covariance corresponding to the peaks and spread of the intensity respectively. $N_f$ and $N_g$ are the total number of multivariate Gaussian distributions in the current and desired intensities, respectively. It is assumed that the desired Gaussian mixture intensity, $\nu_{des}({\bf x},k)$, is known. Eq. \eqref{funcff} includes the summation of the individual birth ($\nu_b({\bf x},k)$), spawn ($\nu_\beta$), and survival ($\nu_{p_s}({\bf x},k)$) Gaussian mixture intensities which simplify to another Gaussian mixture. Note that closed form solutions using Gaussian mixtures exist for cases without the state independent assumption. Additionally, $\sum_{i=1}^{N_f}w_f^{(i)}=N_{\text{total}}(t)$ and $\sum_{i=1}^{N_g}w_g^{(i)}=\bar{N}_{\text{total}}(t)$ where $\bar{N}_{\text{total}}(t)$ is the desired number of agents. The intensity function $\nu({\bf x},t)$ is in terms of the swarm state while $\nu_{des}({\bf x},t)$ is in terms of the desired state.
The swarm intensity function can be propagated through updates on the mean and covariance of the Gaussian mixtures as given by
\begin{equation}\label{dynamicswithu}
{\bf m}_{f,k+1}^i=A_k{\bf m}_{f,k}^i+B_k{\bf u}_{f,k}^i,
\end{equation}
\begin{equation}\label{covary}
P_{f,k+1}^i=A_kP_{f,k}^iA_k^T+Q_k,
\end{equation}
where $Q_k$ is process noise. The agents' states $\mathbf{x}$ are incorporated in the mean and covariance of the Gaussian mixture intensity. Then given the Gaussian mixture intensities assumption, a control variable is calculated for each component ${\bf u}_{f,k}^i$. Additionally, each Gaussian mixture component represents many agents since the intensity function integrates to the total number of agents. Note that although linear dynamics are used, the dynamics can be modeled as a nonlinear function of the state.% with the nonquadratic RFS-based objective function.

Additionally, the measurement update is also closed form given by the intensity
\begin{equation}\label{funcf}
\begin{split}
\nu_k({\bf x},k)=f({\bf x})&=(1-p_d({\bf x}))\bar{\nu}_k({\bf x})\\&+\sum_{{\bf z}\in Z_k}\sum_{j=1}^{N_f}w_k^{(j)}\mathcal{N}\left( \mathbf{x};{\bf m}_{k|k}^{(j)}({\bf z}),P_{k|k}^{(j)}\right),
\end{split}
\end{equation}
where
\begin{subequations}
\begin{equation}
w_k^{(j)}=\frac{p_d({\bf x})w_f^{(j)}q^{(j)}({\bf z})}{K({\bf z})+p_d({\bf x})\sum_{l=1}^{N_f}w_f^{(l)}q^{(l)}({\bf z})},
\end{equation}
\begin{equation}
{\bf m}_{k|k}^{(j)}({\bf z})={\bf m}_{f}^{(j)}+K^{(j)}\left( {\bf z}-H_k{\bf m}_{f}^{(j)}\right),
\end{equation}
\begin{equation}
P_{k|k}^{(j)}=\left( I-K^{(j)}H_k\right)P_f^i,
\end{equation}
\begin{equation}
K^{(j)}=P_f^iH_k^{T}\left( H_kP_f^iH_k^{T}+R_k\right)^{-1},
\end{equation}
\begin{equation}
q_k^{(j)}({\bf z})=\mathcal{N}\left( {\bf z}; H_k{\bf m}_{f}^{(j)},R_k+H_kP_f^iH_k^{T}\right),
\end{equation}
\end{subequations}
which closely follow the Kalman filter measurement update equations.

Each individual swarm agent runs a local PHD observer to estimate the state of the swarm by modeling the swarm as a distribution. Thus, using RFS theory, it is assumed that the individual swarm agents form an intensity function that is a Gaussian mixture intensity in which the means and covariances of the Gaussian mixture are propagated and controlled. An optimal control problem is set up that tracks a desired swarm formation by minimizing its control effort in the following objective function 
%as shown in Equation $\ref{distance}$.
\begin{equation}\label{distance}
J({\bf u})=\int_{0}^T{\bf u}(t)^{T}R{\bf u}(t)+D(\nu({\bf x},t),\nu_{des}({\bf x},t))dt,
\end{equation}
where $\nu_{des}({\bf x},t)$ is the desired formation, $R$ is the positive definite control weight matrix, and ${\bf u}(t)$ is the control effort for the Gaussian mixture intensities shown in Eq. \eqref{dynamicswithu}. Both $\nu({\bf x},t)$ and $\nu_{des}({\bf x},t)$ are defined over the complete state space which include position and velocity parameters. $D(\cdot,\cdot)$ is the distance between Gaussian mixtures which has several closed-form solutions, and it has been used previously to define an objective function for path planning of multi-agent systems \cite{41}. For the Gaussian mixture approximation for RFSs, the objective function is defined by
\begin{equation}\label{overalleqn}
\begin{aligned}
&\min_{{\bf u}_k,k=1,...,T}J({\bf u})=\sum_{k=1}^T\ {\bf u}_k^{T}R{\bf u}_k\\&+\sum_{j=1}^{N_f}\sum_{i=1}^{N_f}w_{f,k}^{(j)}w_{f,k}^{(i)}\mathcal{N}({\bf m}_{f,k}^j;{\bf m}_{f,k}^i,P_{f,k}^i+P_{f,k}^j)\\&+\sum_{j=1}^{N_g}\sum_{i=1}^{N_g}w_{g,k}^{(j)}w_{g,k}^{(i)}\mathcal{N}({\bf m}_{g,k}^j;{\bf m}_{g,k}^i,P_{g,k}^i+P_{g,k}^j)\\&-2\sum_{j=1}^{N_g}\sum_{i=1}^{N_f}w_{g,k}^{(j)}w_{f,k}^{(i)}\mathcal{N}({\bf m}_{g,k}^j;{\bf m}_{f,k}^i,P_{g,k}^i+P_{f,k}^j)\\&- \alpha\sum_{j=1}^{N_g}\sum_{i=1}^{N_f}w_{g,k}^{(j)}w_{f,k}^{(i)}\ln\left(\mathcal{N}({\bf m}_{g,k}^j;{\bf m}_{f,k}^i,P_{g,k}^i+P_{f,k}^j)\right),
\end{aligned}
\end{equation}
\begin{equation}\label{qwer}
\begin{split}
\text{Subject} \hspace{5pt}\text{to}:{\bf m}_{f,k+1}^i=A_k{\bf m}_{f,k}^i+B_k{\bf u}_{f,k}^i,\\
P_{f,k+1}^i=A_kP_{f,k}^iA_k^T+Q_k,
\end{split}
\end{equation}
in discrete time \cite{doerr2019random}. The term ${\bf u}_k=[({\bf u}_{f,k}^1)^T,\cdots,({\bf u}_{f,k}^{N_f})^T]^T$ is the collection of all control variables. Therefore, control solutions are found by either using DDP where the objective function is quadratized by taking a Taylor series approximation about a nominal trajectory or using optimization techniques (e.g. the Quasi-Newton method) where the nonquadratic objective function is used directly to find an optimal control solution.

The key features for the RFS control problem is that it can allow for a unified representation for swarming systems. This unified representation is achieved by minimizing the RFS objective function, Eq. \eqref{overalleqn}, about the swarm intensity statistics given by Eq. \eqref{qwer}. Thus, it can handle multi-fidelity swarm localization and control. The swarm is treated probabilistically and the bulk motion is modeled which allows the theory to handle large numbers of indistinguishable units with unknown swarm size. This reduces the dimensionality of the state while enabling complex behavior. Naturally, the RFS control problem is formulated to enable complex decision making through RFS theory.

\section{Differential Dynamic Programming}
The DDP approach to solving nonlinear and nonquadratic equations uses a second-order approximation of the dynamics and objective function for value iteration. The solution is iterated to improve approximations of the optimal trajectory of the system. \cite{tassa2014control}. Note that if linear dynamics are used, the iterative linear quadratic regulator (ILQR) formulation is obtained \cite{tassa2014control,11}. Since the results are produced by a linear system, both the DDP and ILQR terms can be used interchangeably. For a nonlinear system,
\begin{equation}\label{nonlinsys}
{\bf x}_{k+1}=f({\bf x}_k,{\bf u}_k),
\end{equation}
and a nonquadratic objective function,
\begin{equation}\label{costlqr}
J({\bf x}_0,U)=\sum_{k=0}^{N-1}l({\bf x}_k,{\bf u}_k)+l_f({\bf x}_N),
\end{equation}
where $l({\bf x}_k,{\bf u}_k)$ is the running cost, and $l_f({\bf x}_N)$ is the terminal cost, the cost-to-go is given by
\begin{equation}\label{costtogo}
J({\bf x}_{k},U_{k})=\sum_{k}^{N-1}l({\bf x}_k,{\bf u}_k)+l_f({\bf x}_N),
\end{equation}
starting at state $\mathbf{x}_k$ instead of $\mathbf{x}_0$. The value function, or optimal cost-to-go, can be found by minimizing Eq. \eqref{costtogo} in terms of the control sequence $U_{k}=[{\bf u}_{k},{\bf u}_{k+1},\cdots,{\bf u}_{N-1}]$. By letting $V({\bf x}_{N})=l_f({\bf x}_N)$, the minimization of the control sequence is reduced to a minimization to a control at a time-step by using the principle of optimality, and the value can be solved backwards in time using
\begin{equation}\label{valueupdate}
V({\bf x}_{k})=\min_{\mathbf{u}_{k}} \left( l({\bf x}_k,{\bf u}_k)+V({\bf x}_{k+1})\right).
\end{equation}
With Eq. \eqref{nonlinsys} and Eq.\eqref{valueupdate}, a Taylor series expansion can be found which linearizes and quadratizes the nonlinear system and objective function about perturbations $\delta\mathbf{x}_k$ and $\delta\mathbf{u}_k$ given by
\begin{equation}
\delta\mathbf{x}_{k+1}=f_x\delta\mathbf{x}_k+f_u\delta\mathbf{u}_k,
\end{equation}
\begin{equation}\label{valueexpand}
\begin{split}
Q(\delta\mathbf{x},\delta\mathbf{u})&=l(\mathbf{x}_k+\delta\mathbf{x}_k,\mathbf{u}_k+\delta\mathbf{u})_k-l(\mathbf{x},\mathbf{u})\\&+V({\bf x}_{k+1}+\delta\mathbf{x}_{k+1})-V({\bf x}_{k+1}),\\
&\approx\frac{1}{2}\left[ \begin{array}{c} 1 \\  \delta\mathbf{x}_k \\ \delta\mathbf{u}_k \end{array} \right]^{T} \begin{bmatrix} 0& Q_{x}^{T} & Q_{u}^{T} \\ Q_{x}&Q_{xx} & Q_{xu} \\Q_{u}&Q_{ux} & Q_{uu}  \end{bmatrix}\left[ \begin{array}{c} 1\\  \delta\mathbf{x}_k \\ \delta\mathbf{u}_k \end{array} \right],
\end{split}
\end{equation}
where $f_x$ and $f_u$ are the linearized transition matrices, and $Q_x$, $Q_u$, $Q_{xx}$, $Q_{xu}$, and $Q_{uu}$ are the running weights of the Q-function. The authors would like to note that the time-step is dropped and any primes used denotes the next time-step. The equations for these weights are given by
\begin{subequations}\label{quadapproximations}
\begin{equation}
Q_x= l_x+ f_x^{T}V_x',
\end{equation}
\begin{equation}
Q_u= l_u+ f_u^{T}V_x',
\end{equation}
\begin{equation}
Q_{xx}= l_{xx}+ f_x^{T}V_{xx}'{ f}_x,
\end{equation}
\begin{equation}
Q_{uu}= l_{uu}+ f_u^{T}V_{xx}'{ f}_u,
\end{equation}
\begin{equation}
Q_{ux}= l_{ux}+ f_u^{T}V_{xx}'{ f}_x,
\end{equation}
\end{subequations}
where $l_x$, $l_u$, $l_{xx}$, $l_{uu}$, and $l_{ux}$ are the gradients and Hessians of the cost function and $V_x'$ and $V_{xx}'$ are the gradient and Hessian of the value function. The linear control policy is given by
\begin{subequations}\label{controllerKk}
\begin{equation}
K=-Q_{uu}^{-1}Q_{ux},
\end{equation}
\begin{equation}
\mathbf{k}=-Q_{uu}^{-1}Q_{u},
\end{equation}
\end{subequations}
where $K$ is the local feedback and $\mathbf{k}$ is the feed-forwards gains for the optimal policy, and the gradient and Hessian of the value function have the form
\begin{subequations}\label{vxvxx}
\begin{equation}
\Delta V=-\frac{1}{2}\mathbf{k}^{T}Q_{uu}\mathbf{k},
\end{equation}
\begin{equation}
V_x=Q_x-K^{T}Q_{uu}\mathbf{k},
\end{equation}
\begin{equation}
V_{xx}=Q_{xx}-K^{T}Q_{uu}K.
\end{equation}
\end{subequations}
Therefore, the optimal policy update is given by
\begin{equation}
\mathbf{\hat{u}}_k=\mathbf{u}_k+\mathbf{k}_k+K_k\left(\mathbf{\hat{x}}_k-\mathbf{x}_k\right).
\end{equation}

This algorithm is iterated to continuously obtain better approximations of the optimal trajectory. A tolerance can be set as the cost function converges to its optimal trajectory to end the iteration.

\section{Decentralized Control Formulation}
%Subsection text here.

%\IEEEpubidadjcol
The framework for decentralizing RFS control for swarming agents is to design sparse control matrices using sparse LQR \cite{fardad2011sparsity,lin2012sparse,lin2013design}. The following discussion on sparse LQR follows closely to Lin's work on sparse feedback gains \cite{lin2012sparse}.
\subsection{Sparse LQR Problem}
The continuous state-space representation of a linear time-invariant dynamical system with a structured control design is represented by
\begin{subequations}\label{coneqnuu}
\begin{equation}\label{coneqn}
    \dot{\mathbf{x}}(t)=A_c\mathbf{x}(t)+B_c\mathbf{u}(t)+B_{c2}\mathbf{d}(t),
\end{equation}
\begin{equation}\label{coneqnu}
    \mathbf{u}(t)=-F\mathbf{x}(t),
\end{equation}
\end{subequations}
where $A_c$ is a continuous state transition matrix, $B_c$ is a continuous control transition matrix, $\mathbf{d}(t)$ is a disturbance or external input for a time $t$, $B_{c2}$ is the disturbance transition matrix, and $F$ is a state feedback (control) gain dependent on the sparsity (structural) constraints $F\in \mathcal{S}$. A sparsity constraint subspace $\mathcal{S}$ is assumed to be non-empty for all sparsity patterns for controller gains that are stable. For an infinite horizon LQR, the total cost is quadratic in terms of the state and control given by
\begin{equation}\label{conquadcost}
    J(\mathbf{x}(t),\mathbf{u}(t))=\int_{0}^{\infty}\mathbf{x}(t)^T Q\mathbf{x}(t)+\mathbf{u}(t)^T R\mathbf{u}(t),
\end{equation}
where $Q$ is a positive semi-definite state weight matrix and $R$ is a positive definite control weight matrix. By plugging in Eq. \eqref{coneqnuu} into Eq. \eqref{conquadcost} for control gain $F$ \cite{levine1970determination}, the optimal control problem with structural constraints becomes
\begin{equation}\label{objF}
\resizebox{0.48\textwidth}{!}{$
\begin{aligned}
    &\min J(F)=\text{trace}\left( B_{c2}^T \int_0^{\infty}\text{e}^{\left(A-B_cF\right)^Tt}\left( Q+F^TRF\right)\text{e}^{  \left( A-B_cF \right) t}dtB_{c2}\right)\\
    &\text{Subject to:}\hspace{12pt} F\in \mathcal{S}.
\end{aligned}$}
\end{equation}
The objective is to determine a control gain, $F\in\mathcal{S}$, that minimizes the LQR cost. Fortunately, the integral in Eq. \eqref{objF} is bounded for stabilizing $F$, thus a control solution can be found using the Lyapunov equation given by
\begin{equation}
    (A-B_cF)^TP+P(A-B_cF)=-(Q+F^TRF),
\end{equation}
which reduces the $J(F)$ into
\begin{equation}
    J(F)=\text{trace}\left( B_{c2}^TP(F)B_{c2}\right).
\end{equation}
The control objective in Eq. \eqref{objF} assumes the sparsity constraints are known before the optimization takes place, but these constraints may be unknown and appropriate sparsity patterns for decentralized control must be found. The optimization problem can be modified to provide a sparsity promoting optimal control solution which provides the performance and the topology for decentralized control. The sparsity-promoting optimal control problem is
\begin{subequations}\label{sparseopt}
\begin{equation}
    \min J(F)+\gamma g_0(F),
\end{equation}
\begin{equation}\label{nnzf}
     g_0(F)=\textbf{nnz}(F),
\end{equation}
\end{subequations}
where $g_0(F)$ is the number of non-zeros ($\mathbf{nnz(\cdot)}$) for control gain $F$ and $\gamma\geq0$ is a scalar weight to penalize $g_0(F)$. By including the number of non-zeros in the control gain $F$ into the control objective directly, sparsity in $F$ is directly promoted in the optimization of the problem. More zeros (sparsity) in a control gain matrix corresponds to more localization in the information topology network. The weight $\gamma$ follows similarly to how the $Q$ and $R$ matrices penalize $\mathbf{x}$ and $\mathbf{u}$, respectively, but $\gamma$ penalizes the number of non-zeros in $F$. For example, when $\gamma>>0$, the number of non-zeros in $F$ is penalized heavily, thus, $\gamma$ promotes more localized control. When $\gamma=0$, no penalization of the control gain takes place, and a standard LQR solution with a centralized control gain matrix is found.
\subsection{Sparsity-Promoting Optimal Control}
The function $g_0(F)$ is a nonconvex argument in the optimization problem. As a result, finding the solution involves a brute-force search which becomes intractable. To circumvent this issue, the $g_0(F)$ function is substituted with the $L_1$ norm which is a nondifferentiable convex function given by
\begin{equation}\label{gl1}
    g_1(F)=||F||_1=\sum_{i,j}|F_{ij}|,
\end{equation}
which gives higher costs to non-zeros elements in $F$ with larger magnitudes \cite{boyd2004convex}. This differs from $g_0(F)$ which gives the same cost to all non-zero elements. Therefore, the $L_1$ norm becomes a convex relaxation of the original problem, but the original $g_0(F)$ can be approximated better or recovered exactly by using a weighted $L_1$ norm given by
\begin{equation}\label{gwl1}
    g_2(F)=\sum_{i,j}W_{ij}|F_{ij}|,
\end{equation}
where $W_{ij}$ are positive weights. The weights can be used to approximate the $L_1$ norm closer to $g_0(F)$, but if $W_{ij}$ is chosen to be inversely proportional to $|F_{ij}|$ as given by
\begin{equation}\label{nnzlw1}
    W_{ij}=
    \begin{cases}
            1/|F_{ij}|, &\text{if }         F_{ij}\neq 0,\\
            \infty, &   \text{if }      F_{ij}=0,
    \end{cases}
\end{equation}
the weighted $L_1$ norm and $g_0(F)$ equate to
\begin{equation}
    \sum_{i,j}W_{ij}|F_{ij}|=\textbf{nnz}(F).
\end{equation}
Although the weighted $L_1$ norm is viable to recover $g_0(F)$, the weights are dependent on the unknown feedback gain $F$. Therefore, an iterative algorithm, the alternating direction method of multipliers (ADMM), is used which trades off optimal performance, $J$, and sparsity, $\gamma$. First, initial centralized control gain, $F$, with $\gamma=0$ is inputted into ADMM. Then, $\gamma$ is increased and the ADMM iterative algorithm is used in conjunction with $F$ and the previous $\gamma$ to obtain a sparser $F$. Once the desired sparsity is found, the sparsity structure is fixed and the sparse control gain is found using the structured optimal control problem in Eq. \eqref{objF}. The method by which sparsity structures are identified using ADMM is explained in the next discussion.
\subsection{Alternating Direction Method of Multipliers}
The optimization problem in Eq. \eqref{sparseopt} can be rearranged into a constrained optimization problem
\begin{equation}\label{objFsplit}
\begin{split}
   &\min J(F)+\gamma g(G),\\
    &\text{Subject to:}\hspace{12pt} F-G=0, 
\end{split}
\end{equation}
where $G$ decouples the sparsity cost separately from the performance cost. The equality constraint $F-G=0$ makes Eq. \eqref{objFsplit} equivalent to Eq. \eqref{objF}. The associated augmented Lagrangian to the constrained optimization problem is 
\begin{equation}\label{auglag}
\begin{split}
\mathcal{L}_{\rho}(F,G,\Lambda)&=J(F)+\gamma g(G)\\
&+\text{trace}(\Lambda^T(F-G))+\frac{\rho}{2}||F-G||^2_F,
\end{split}
\end{equation}
where $\lambda$ is the Langrange multiplier, $\rho>0$ is scalar, and $||\cdot||_F$ is the Frobenius norm. By decoupling $J$ and $g$, the structures for both $J$ and $g$ can be exploited using the ADMM algorithm optimization. The ADMM algorithm contains the $F$-minimization, $G$-minimization, and Lagrange multiplier steps in which $F$ and $G$ are minimized iteratively \cite{boyd2011distributed}. This is given by
\begin{subequations}
\begin{equation}\label{fminopt}
    F^{k+1}=\text{arg}\min_F\mathcal{L}_{\rho}(F,G^k,\Lambda^k),
\end{equation}
\begin{equation}\label{gminopt}
    G^{k+1}=\text{arg}\min_G\mathcal{L}_{\rho}(F^{k+1},G,\Lambda^k),
\end{equation}
\begin{equation}
    \Lambda^{k+1}=\Lambda^k+\rho(F^{k+1}-G^{k+1}),
\end{equation}
\end{subequations}
and the convergence tolerance
\begin{equation}
    ||F^{k+1}-G^{k+1}||_F\leq\epsilon \hspace{12pt} \text{and} \hspace{12pt} ||G^{k+1}-G^{k}||_F\leq\epsilon.
\end{equation}
The $F$-minimization and $G$-minimization alternate direction in terms of finding the optimal $F$ and $G$, respectively, which gives ADMM its namesake. The Lagrange multiplier update steps with a size $\rho$ which guarantees the feasibility of finding $G^{k+1}$ and $\Lambda^{k+1}$.

For the sparsity-promoting optimization problem, ADMM provides benefits in the separability and differentiability of the sparsity cost and the performance cost. When calculating the performance cost using the control gain matrix, the matrix cannot be separated into individual elements to find optimal solutions. By separating optimization in the $F$-minimization and $G$-minimization steps, the G-minimization step can be decomposed into subproblems that involve individual elements (scalars) of the control gain matrix. Therefore, an optimal solution can be found analytically using either $g_0(F)$, $g_1(F)$, or $g_2(F)$. The other benefit to ADMM is differentiability. The performance cost is differentiable in terms of the control gain, but the sparsity cost is non-differentiable as discussed before. By separating the optimization problem in two steps, gradient descent algorithms can be used for the $F$-minimization step, and analytical solutions can be found for the $G$-minimization step. This is discussed next.
\subsubsection{The $F$-Minimization Step Solution}
The minimization of Eq. \eqref{fminopt} can use any descent method. Although gradient descent or Newton's methods can be used, the Anderson-Moore descent can converge faster than gradient descent and is simpler to implement than Newton's method \cite{makila1987computational}. From the augmented Lagrangian in Eq. \eqref{auglag}, an equivalent optimization problem can be obtained by completing the square given by
\begin{equation}
\begin{gathered}
    \min \phi(F)=J(F)+(\rho/2)||F-U^k||^2_F,\\
    U^k=G^k-(1/\rho)\Lambda^k.
    \end{gathered}
\end{equation}
Using methods developed in \cite{levine1970determination,rautert1997computational}, the necessary conditions for optimality are obtained as
\begin{subequations}
\begin{equation}\label{ctrbgram}
 (A-B_cF)L+L(A-B_cF)^T=-B_{c2}B_{c2}^T,   
\end{equation}
\begin{equation}\label{obsvgram}
  (A-B_cF)^TP+P(A-B_cF)=-(Q+F^TRF),     
\end{equation}
\begin{equation}\label{optimalitysol}
  \nabla\phi(F)=2RFL+\rho F-2B_c^TPL-\rho U^k=0,  
\end{equation}
\end{subequations}
where Eqs. \eqref{ctrbgram} and \eqref{obsvgram} are the controllability and observability grammians, respectively, and Eq. \eqref{optimalitysol} is the optimality condition for $\mathcal{L}_p$. Anderson-Moore iteratively solves for Eqs. \eqref{ctrbgram} and \eqref{obsvgram} for $L$ and $P$ with a fixed $F$ using the solution to the Lyapunov equations, and then solves $F$ in Eq. \eqref{optimalitysol} with a fixed $L$ and $P$ using the solution to the Sylvester equation to obtain a new $\bar{F}$ \cite{makila1987computational,lin2012sparse}. This consists of one iteration for the F-minimization step. To complete the F-minimization step, a descent direction, $\tilde{F}=\bar{F}-F$, is obtained to allow for convergence to a stationary point on $\phi$. The stationary point $\phi$ is locally convex and provides a local minimum on $\phi$. Note that step-size rules (i.e. determining $s$ in $F+s\tilde{F}$ using the Armijo rule) can be used to guarantee convergence to the stationary point \cite{bertsekas2006nonlinear}.
\subsubsection{The $G$-Minimization Step Solution}
To find an analytical solution to the $G$-minimization in Eq. \eqref{gminopt}, the first step is to complete the square of Eq. \eqref{auglag} with respect to $G$. This is given by
\begin{equation}
    \begin{gathered}
    \min \phi(G)=\gamma g(G)+(\rho/2)||G-V^k||^2_F,\\
    V^k=(1/\rho)\Lambda^k+F^{k+1}.
    \end{gathered}
\end{equation}
This equation can be reduced into summation of element-wise components (scalars) by substituting the $g(\cdot)$ functions from Eqs. \eqref{nnzf}, \eqref{gl1}, or \eqref{gwl1} and solving directly. The weighted $L_1$ , Eq. \eqref{gwl1}, is a general function for Eq. \eqref{gl1} when $W_{ij}=1$ and Eq. \eqref{nnzf} when Eq. \eqref{nnzlw1}, so the objective can be reduced element-wise using a strictly convex Eq. \eqref{gwl1} given by
\begin{equation}
    \phi(G)=\sum_{i,j}\left( \gamma W_{ij}|G_{ij}|+(\rho/2)(G_{ij}-V_{ij}^k)^2\right).
\end{equation}
Thus, the minimization is
\begin{equation}
    \min \phi_{ij}(G_{ij})=\gamma W_{ij}|G_{ij}|+(\rho/2)(G_{ij}-V_{ij}^k)^2,
\end{equation}
for each element in $G$. The unique solution to this problem is 
\begin{equation}
    G_{ij}^*=
        \begin{cases}
            V_{ij}^k-a, &         V_{ij}^k\in(a,\infty),\\
            0, &         V_{ij}^k\in(-a,a),\\
            V_{ij}^k+a, &         V_{ij}^k\in(-\infty,-a),
    \end{cases}
\end{equation}
where $a=(\gamma/\rho)W_{ij}$ is a scalar. This equation is the shrinkage operator \cite{boyd2011distributed}, and it is the solution when Eq. \eqref{gl1} or \eqref{gwl1} is substituted. The amount by which $ G_{ij}^*$ is minimized is the parameter $a$. If $\gamma$ or $W_{ij}$ is increased, the minimization becomes more forceful. This occurs similarly by reducing $\rho$.
If Eq. \eqref{nnzf} is used, the $G$-minimization reduces to
\begin{equation}
     \min \phi_{ij}(G_{ij})=\gamma\textbf{nnz}(G_{ij})+(\rho/2)(G_{ij}-V_{ij}^k)^2,
\end{equation}
and has a unique solution given by
\begin{equation}
G_{ij}^*=
        \begin{cases}
            0, &         |V_{ij}^k|\leq b,\\
            V_{ij}^k, &         |V_{ij}^k|> b,
    \end{cases}
\end{equation}
where $b=\sqrt{2\gamma/\rho}$ is a scalar. This is the truncation operator \cite{lin2013design}. By using any of the $g(\cdot)$ functions, a unique solution for the optimization in Eq. \eqref{gminopt} can be found.

\section{Application to RFS Control}
The theory for using Sparse LQR for decentralized control is formulated in a continuous time representation given by Eqs. \eqref{coneqnuu} and \eqref{conquadcost}. Unfortunately, RFS control is formulated in discrete time with a zero-order hold on control. Therefore, a bridge between the two theories must be found. Previously, sparse feedback gains have been found in discrete time using non-convex sparsity-promoting penalty functions using sequential convex optimization \cite{fardad2014design}, but for this work, a less computationally intensive and theoretically extensive method is more useful. Work in discretizing the sparse LQR formulation has been made by high level discussion of using discrete Lyapunov and Sylvester equations, although no theory or algorithms have been presented \cite{verdoljak2016application}. This method is used to obtain a discrete version of sparse LQR which is used directly with the centralized gain outputted by RFS control. Thus, the output of sparse LQR is a decentralized RFS gain which contains an information topology that is localized or fully decentralized. 

\section{Relative Motion using Clohessy-Wiltshire Equations}\label{dynamicmodelsec}
To show viability of decentralized swarm control via RFS, a spacecraft relative motion model described by a linear system is used to model satellite dynamics. The dynamic equations of individual agents are used to describe the dynamics of the Gaussian mixture components (means) given by the control objective Eqs. \eqref{overalleqn} and \eqref{qwer}. Since linear dynamics are used, the DDP term can be expressed as ILQR.

For a spacecraft in low Earth orbit, the relative dynamics of each spacecraft (agent), to a chief spacecraft in circular orbit, is given by the Clohessy-Wiltshire equations \cite{42}
\begin{subequations}
\begin{equation}
\ddot{x}=3n^2x+2n\dot{y}+a_x,
\end{equation}
\begin{equation}
\ddot{y}=-2n\dot{x}+a_y,
\end{equation}
\begin{equation}
\ddot{z}=-n^2z+a_z,
\end{equation}
\end{subequations}
where $x$, $y$, and $z$ are the relative positions in the orbital local-vertical local-horizontal (LVLH) frame and $a_x$, $a_y$, and $a_z$ are the accelerations in each axis respectively. The variable $n$ is defined as the orbital frequency given by
\begin{equation}
n=\sqrt{\frac{\mu}{a^3}},
\end{equation}
where $\mu$ is the standard gravitational parameter and $a$ is the radius of the circular orbit. The continuous state-space representation is given by
\begin{equation} \label{qqMatrix}
 A_c = \begin{bmatrix}0&0& 0 & 1 &0&0\\ 0 & 0&0&0&1&0\\0&0&0&0&0&1\\3n^2&0&0&0&2n&0\\0&0&0&-2n&0&0\\0&0&-n^2&0&0&0 \end{bmatrix}, \hspace{6pt} B_c = \begin{bmatrix} 0&0&0 \\0&0&0\\0&0&0\\1&0&0\\0&1&0\\0&0&1  \end{bmatrix},
\end{equation}
with a state vector $\mathbf{x}=\left[ x, y, z, \dot{x}, \dot{y}, \dot{z}\right]$ and a control input $\mathbf{u}=\left[ a_x, a_y, a_z\right]^{T}$. The $A_c$ and $B_c$ matrices are discretized along a fixed time interval utilizing a zero-order hold assumption for the control (i.e. control is held constant over the time-interval). This results in discretized $A$ and $B$ matrices for the state-space equation,
\begin{equation}\label{discretization}
{\bf x}_{k+1}=A{\bf x}_{k}+B{\bf u}_{k}.
\end{equation}

\section{Results}
Decentralized RFS control is implemented using the relative motion Clohessy-Wiltshire dynamics with different sparsity ($\gamma$)~weights. Specifically, RFS control is implemented using the $L_2^2$ plus quadratic distance and ILQR. The dynamics model for agents within the swarm are decoupled from each other, but the distributional distance-based cost may have coupling between agents. Therefore, an RFS control gain that is found will be centralized due to coupling in the objective function. Then, the control gain matrix is decentralized by varying the $\gamma$ parameter and using sparse LQR. Three cases with varying $\gamma$ are implemented to show how changes in information topology affect performance of the agents in action.
\subsection{Case 1: Centralized Control}
\begin{figure}
\begin{centering}
    \subfigure[Trajectory Snapshots]{
      \includegraphics[keepaspectratio,trim={0 .15cm 0 .27cm},clip,width=.4\textwidth]{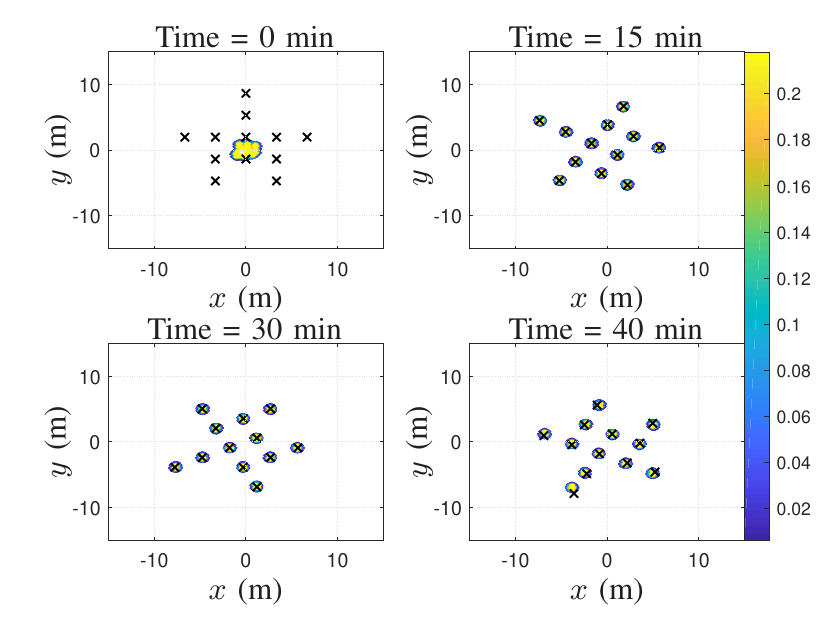}
%      \psfragfig*[width=.46\textwidth]{figures/sparsetraj1}{
%      \psfrag{time = 0 min}[][]{\small{$x$ (m)}}
%      \psfrag{time = 15 min}[][]{\small{$x$ (m)}}
%      \psfrag{time = 30 min}[][]{\small{$x$ (m)}}
%      \psfrag{time = 40 min}[][]{\small{$x$ (m)}}
%      \psfrag{x}[][]{\small{$x$ (m)}}
%      \psfrag{y}[][]{\small{$y$ (m)}}}
      \label{fig:traj1}}
          \subfigure[Number of Non-Zeros in $K$]{
    \includegraphics[keepaspectratio,trim={0 1.0cm 0 1.25cm},clip,width=.4\textwidth]{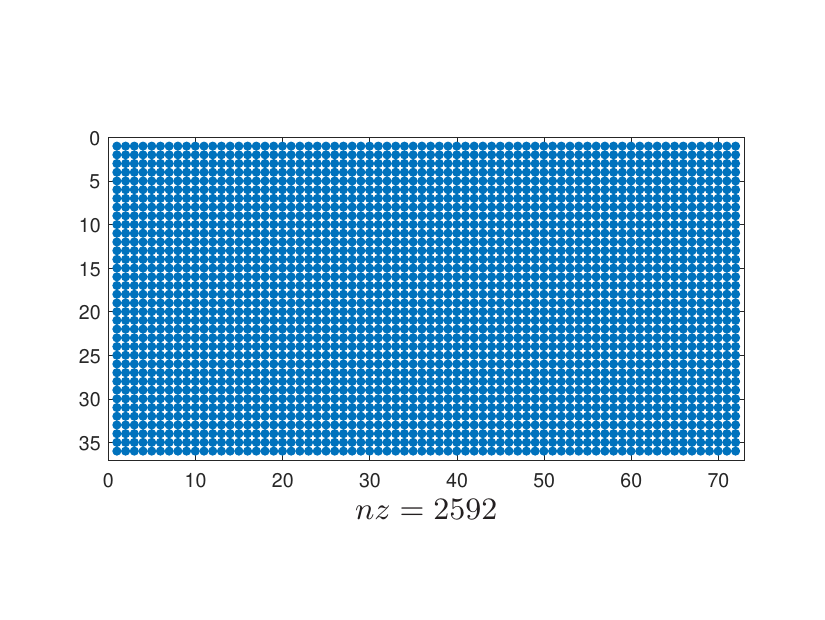}
%    \psfragfig*[width=.46\textwidth]{figures/sparsennz1}{
%      \psfrag{nz = 2592}[][]{\small{$nz=2592$}}}
    \label{fig:nnz1}} 
    \caption{Trajectory and number of non-zeros of control gain $K$ for the centralized RFS control case.}\label{fig:centcase}
\end{centering}
\end{figure}

For Case 1, 12 swarm Gaussian mixtures are birthed at the initial time (described by the contours) from uniformly random initial conditions between -1 and 1 m from a chief satellite in a circular orbit. A $\gamma=0$ is applied to the problem which provides no penalty in the sparsity-promoting objective. %This setup is similar to the Relative Motion with Perfect Information in \ref{relativemotionperfinfo} but this example uses a 12 agent swarm instead. 
Figure \ref{fig:traj1} shows the trajectory snapshots of the spacecraft (contours) and the desired Gaussian mixtures (black x's) using the aforementioned $L_2^2$ plus quadratic divergence and ILQR control. Through time, the swarm intensity converges quickly into the rotating star-shaped formation (where each contour contains a single agent and its covariance) and maintains the formation for a duration of 40 min. Figure \ref{fig:nnz1} shows the number of non-zeros in the control gain $K$. The control gain matrix of a single agent under Clohessy-Wiltshire dynamics is size $3\times 6$. Therefore, the size for the control gain $K$ of the entire 12 agent swarm is $(3\cdot12)\times (6\cdot12)$. With $\gamma=0$, Figure \ref{fig:nnz1} has no elements that are zero. Each sub-block that contains the $3\times 6$ sized matrix is non-zero which totals to 2592 non-zero elements in $K$. Therefore, each agent in the swarm requires some control information from all the other agents in the field to take an action. Figure \ref{fig:node1} shows the information graph between all the agents. Every agent in the field requires a signal to take an action, but the signals from agents further away from each other may provide a minimal control performance boost in terms of computational power needed. Thus, the control-gain is sparsified to reduce the complexity of the entire network to take an action.
\subsection{Case 2: Localized Control}
\begin{figure}
\begin{centering}
    \subfigure[Trajectory Snapshots]{
      \includegraphics[keepaspectratio,trim={0 .15cm 0 .27cm},clip,width=.4\textwidth]{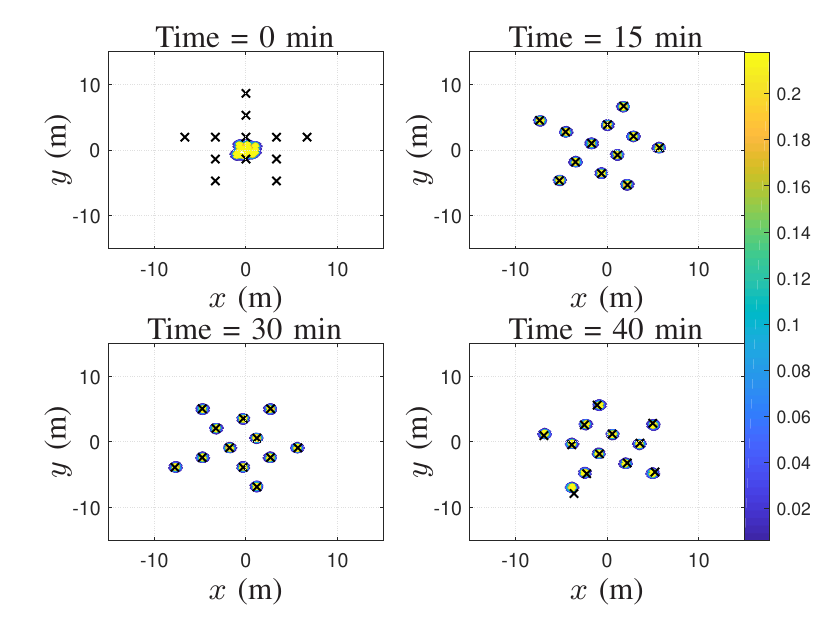}
%      \psfragfig*[width=.46\textwidth]{figures/sparsetraj2}{
%      \psfrag{time = 0 min}[][]{\small{$x$ (m)}}
%      \psfrag{time = 15 min}[][]{\small{$x$ (m)}}
%      \psfrag{time = 30 min}[][]{\small{$x$ (m)}}
%      \psfrag{time = 40 min}[][]{\small{$x$ (m)}}
%      \psfrag{x}[][]{\small{$x$ (m)}}
%      \psfrag{y}[][]{\small{$y$ (m)}}}
      \label{fig:traj2}}
          \subfigure[Number of Non-Zeros in $K$]{
    \includegraphics[keepaspectratio,trim={0 1.0cm 0 1.25cm},clip,width=.4\textwidth]{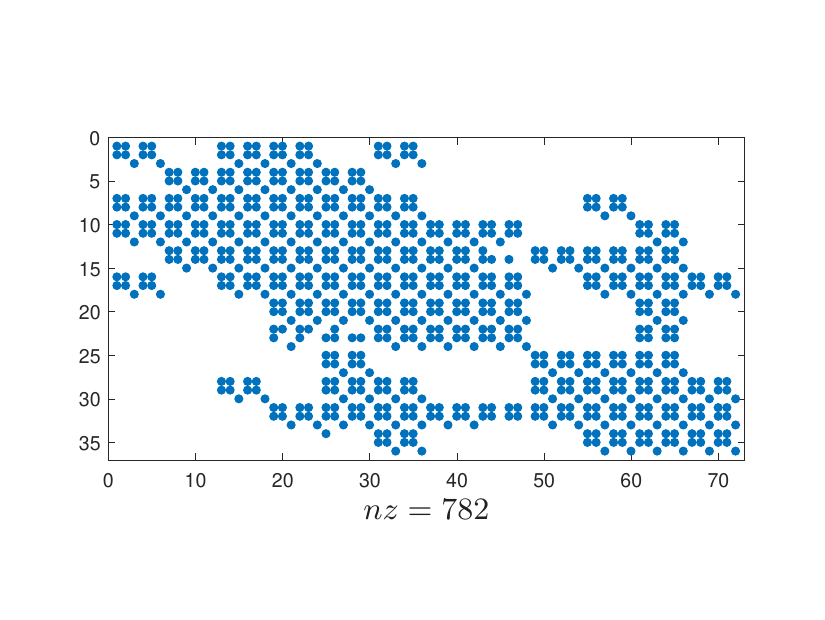}
%    \psfragfig*[width=.46\textwidth]{figures/sparsennz2}{
%      \psfrag{nz = 529}[][]{\small{$nz=529$}}}
    \label{fig:nnz2}} 
    \caption{Trajectory and number of non-zeros of control gain $K$ for the localized RFS control case.}\label{fig:localcase}
\end{centering}
\end{figure}

Case 2 illustrates the effect of promoting sparsity with a $\gamma=10^{-19}$ for the same 12 agent problem. With $\gamma=10^{-19}$, the number of non-zeros is penalized in the sparsity-promoting function in Eq. \eqref{sparseopt}. Figure \ref{fig:traj2} shows the trajectory snapshots of the swarm using localized RFS control. Through time, the swarm intensity converges almost as quickly into the rotating star-shaped formation for the 40 min duration. Specifically from Table \ref{tab:svp}, there is only a reduction of $6\times 10^{-8}\%$ in performance in terms of the centralized performance, the $J_c$ cost, due to localizing the control. Figure \ref{fig:nnz2} shows the number of non-zeros in the control gain $K$. From the figure, the number of non-zeros is reduced to 782 which is $30.2\%$ of the number of non-zeros from the centralized gain, $K_c$, case in Table \ref{tab:svp}. From Figures \ref{fig:nnz2} and \ref{fig:node2}, the agents use the control information from agents local to it. As the control gain matrix becomes more localized, the number of non-zeros in $K$ become increasingly diagonalized with a smaller spread. This is the inherent nature in decentralizing control using sparse LQR. The sparsity-promoting penalty function allows for reduction in the control information needed from individual agents to provide stable localized control with minimal effects on performance.
\subsection{Case 3: Fully Decentralized Control}
\begin{figure}
\begin{centering}
    \subfigure[Trajectory Snapshots]{
      \includegraphics[keepaspectratio,trim={0 .15cm 0 .27cm},clip,width=.4\textwidth]{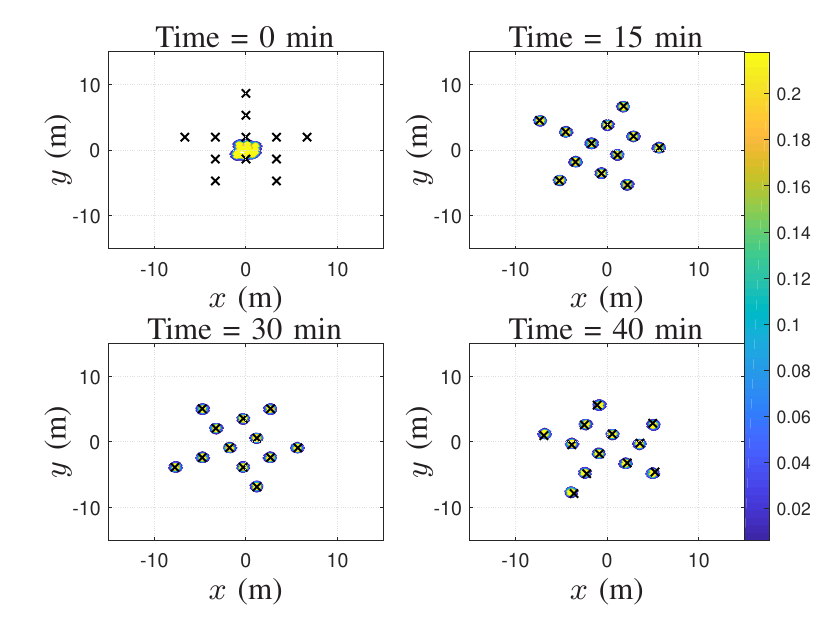}
%      \psfragfig*[width=.46\textwidth]{figures/sparsetraj3}{
%      \psfrag{time = 0 min}[][]{\small{$x$ (m)}}
%      \psfrag{time = 15 min}[][]{\small{$x$ (m)}}
%      \psfrag{time = 30 min}[][]{\small{$x$ (m)}}
%      \psfrag{time = 40 min}[][]{\small{$x$ (m)}}
%      \psfrag{x}[][]{\small{$x$ (m)}}
%      \psfrag{y}[][]{\small{$y$ (m)}}}
      \label{fig:traj3}}
          \subfigure[Number of Non-Zeros in $K$]{
    \includegraphics[keepaspectratio,trim={0 1.0cm 0 1.25cm},clip,width=.4\textwidth]{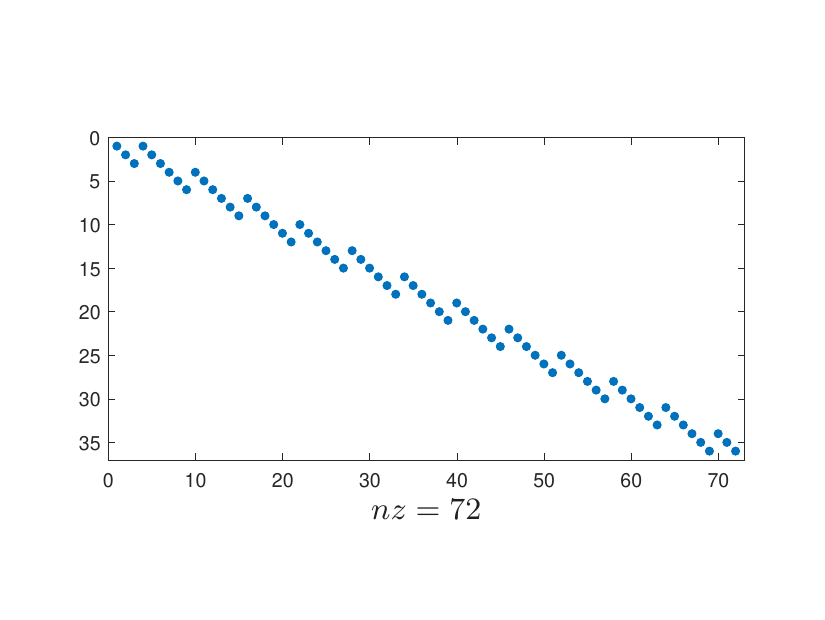}
%    \psfragfig*[width=.46\textwidth]{figures/sparsennz3}{
%      \psfrag{nz = 72}[][]{\small{$nz=72$}}}
    \label{fig:nnz3}} 
    \caption{Trajectory and number of non-zeros of control gain $K$ for the decentralized RFS control case.}\label{fig:decentcase}
\end{centering}
\end{figure}

\begin{figure}[!htb]
\begin{centering}
    \subfigure[Centralized]{
\includegraphics[keepaspectratio,trim={1.25cm .05cm 1.45cm .42cm},clip,width=.28\textwidth]{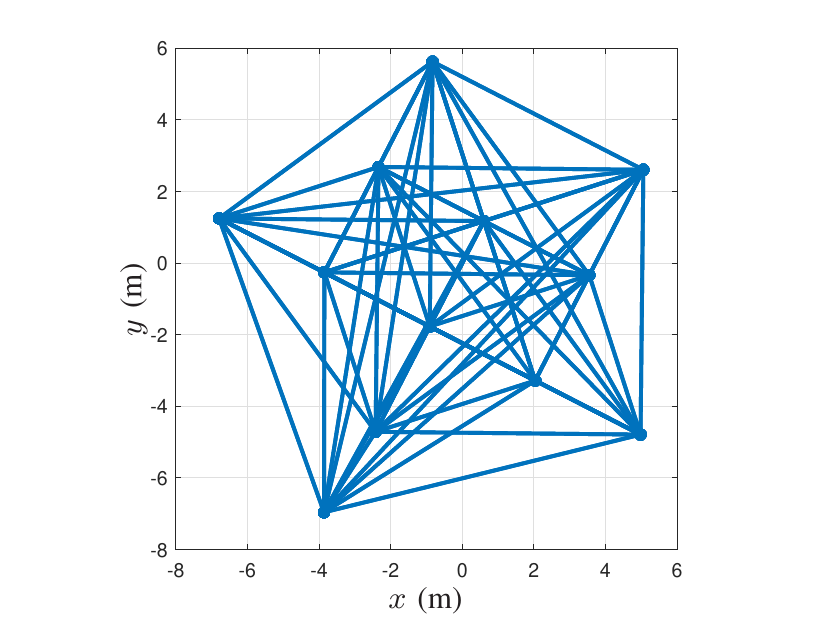}
%      \psfragfig*[width=.30\textwidth]{figures/sparsenode1}{
%      \psfrag{x (m)}[][]{\small{$x$ (m)}}
%      \psfrag{y (m)}[][]{\small{$y$ (m)}}}
      \label{fig:node1}}
\subfigure[Localized]{
\includegraphics[keepaspectratio,trim={1.25cm .05cm 1.45cm .42cm},clip,width=.28\textwidth]{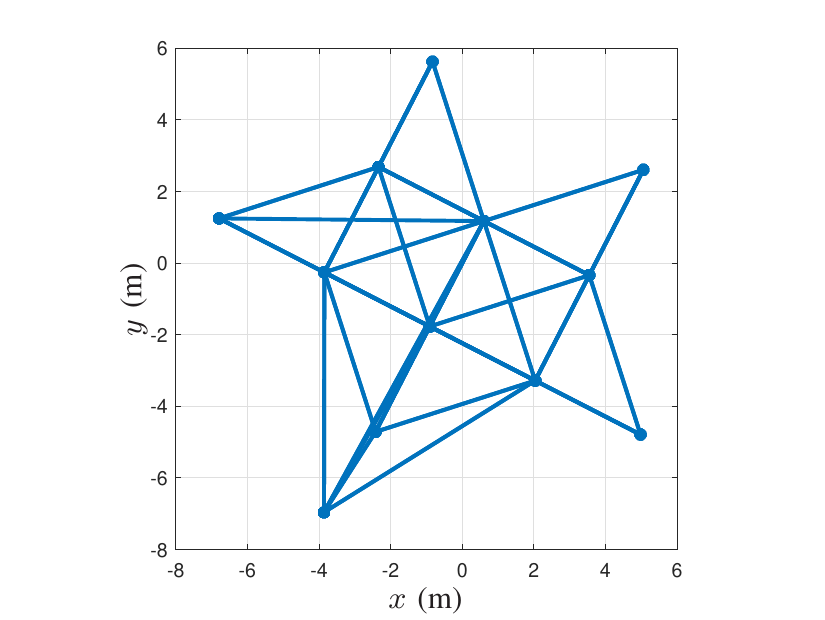}
%    \psfragfig*[width=.30\textwidth]{figures/sparsenode2}{
%      \psfrag{x (m)}[][]{\small{$x$ (m)}}
%      \psfrag{y (m)}[][]{\small{$y$ (m)}}}
    \label{fig:node2}} 
\subfigure[Decentralized]{
\includegraphics[keepaspectratio,trim={1.25cm .05cm 1.45cm .42cm},clip,width=.28\textwidth]{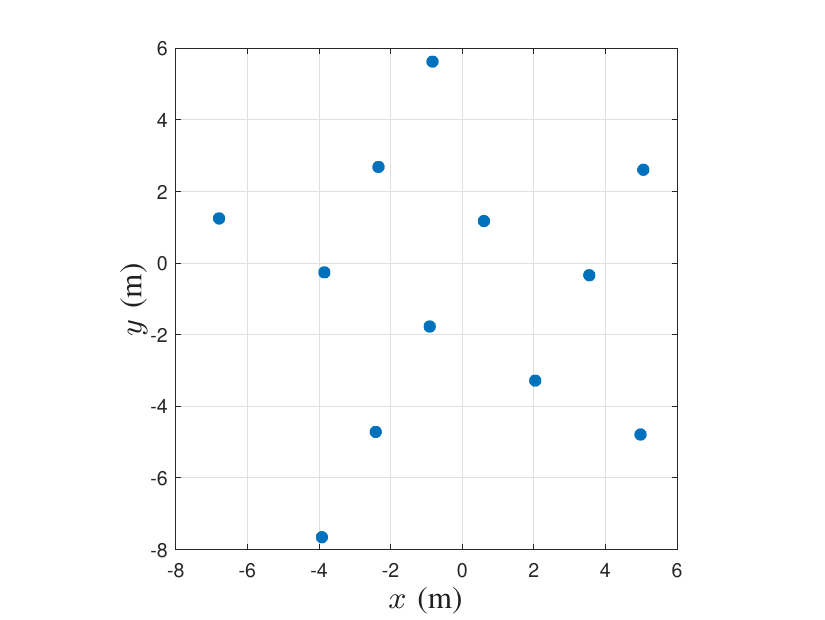}
%      \psfragfig*[width=.30\textwidth]{figures/sparsenode3}{
%      \psfrag{x (m)}[][]{\small{$x$ (m)}}
%      \psfrag{y (m)}[][]{\small{$y$ (m)}}}
      \label{fig:node3}} 
        \caption{Information graph of the 12 agent swarm for $\gamma=$ $0$, $10^{-19}$, and $0.7$ for Figures \ref{fig:node1}, \ref{fig:node2}, and \ref{fig:node3}, respectively.}\label{fig:nodes}
\end{centering}
\end{figure}

Case 3 shows the effect of promoting sparsity with a larger penalty, $\gamma=0.7$, for the 12 agent problem. Figure \ref{fig:traj3} shows the trajectory snapshots of the swarm moving in a fully decentralized manner using RFS control. The swarm intensity converges into the rotating star-shaped formation for the 40 min duration. Performance-wise, there is only a $0.4\%$ reduction in performance compared to $\gamma=0$ example in Table \ref{tab:svp}. Figure \ref{fig:nnz3} shows the number of non-zeros in the control gain $K$. The total number of non-zeros in $K$ is 72 which is $2.8\%$ of $\gamma=0$ in Table \ref{tab:svp}. In this case, the $3\times 6$ sub-matrices occur directly across the diagonal with no spread. No control information is collected from other agents in the swarm which is observed in Figure \ref{fig:node3}. Increasing the $\gamma$ weight penalizes the number of non-zeros in $K$ which allows for more localized, and in this case, a fully decentralized RFS control.

\begin{table}[hbt!]
\caption{\label{tab:svp} Sparsity vs. Performance for Swarm System}
\centering
\begin{tabular}{lcr}
\hline
&Localized &Decentralized \\ \hline
$\textbf{nnz}(K)/\textbf{nnz}(K_c)$ & $30.2\% $ & $2.8\%$ \\ \hline
    $(J-J_c)/J_c$ & $6\times 10^{-8}\%$ & $0.4\% $ \\ \hline
\hline
\end{tabular}
\end{table}

\section{Conclusion}
The objective of this work is to formulate the multi-target estimation and control background for large collaborative swarms using RFS and decentralizing the information topology for control by considering sparse control gain matrices. To provide a control topology that is localized or decentralized, sparse control gain matrices are obtained by sparsifying the RFS control gain matrix using sparse LQR. This allows agents to use local information topology or fully decentralized topology to drive an agent to a target. Specifically by decentralizing the RFS control gain, there is only minimal performance reduction when compared to the centralized gain while reducing the control information necessary for an agent to take an action. Thus, decentralized RFS control becomes more tangible to scientific exploration, communication relaying, self-assembly, and surveillance by allowing agents to use localized information to meet a control objective.

% or


%\appendices

%\section{}

%\section*{Acknowledgment}


\ifCLASSOPTIONcaptionsoff
  \newpage
\fi



\bibliographystyle{IEEEtran}
\bibliography{IEEEabrv,IEEEexample}
\end{document}